# Temperature dependence of the visibility in an electronic Mach-Zehnder interferometer

Masayuki Hashisaka[a)*†], Andreas Helzel [b)†], Shuji Nakamura [a)], Leonid Litvin [b)], Yoshiaki Yamauchi[a)], Kensuke Kobayashi[a)], Teruo Ono[a)], Hans-Peter Tranitz [b)], Werner Wegscheider [b)], and Christoph Strunk [b)].

[a)] *Institute for Chemical Research, Kyoto University, Uji, Kyoto, 611-0011, Japan*

[b)] *Institute für experimentalle und angewandte Physik, Universität Regensburg, Germany*

[*]Corresponding author; TEL/FAX: +81-3-5734-2809; Email: Hashisaka@phys.titech.ac.jp; Present address: Department of Physics, Tokyo Institute of Technology, 2-12-1-H81, Meguro, Tokyo, 152-8551, Japan.

[†] These authors contributed equally to this work.

**Abstract**

We performed the conductance and the shot noise measurements in an electronic Mach-Zehnder interferometer. The visibility of the interference is investigated as a function of the electron temperature that is derived from the thermal noise of the interferometer. The non-equilibrium noise displays both *h/e* and *h/2e* oscillations vs. the modulation gate voltage.

**Keywords**

Mach-Zehnder interferometer; quantum interference; electron temperature

**PACS codes:** 73.23.Ad, 72.70.+m, 73.50.Lw, 71.10.Pm

# 1. Introduction

Electron coherence in solids, which has been one of the central issues in solid state physics, is now attracting renewed attention with the growing expectation for future quantum information technologies. The electronic Mach-Zehnder interferometer (MZI), which was recently realized experimentally [1-4], is a promising device to promote our understanding on the electron coherence [1-7] and to realize the orbital entanglement of electrons [8,9]. The origin of dephasing of electrons has been already addressed in several experiments by using the MZI; the controlled dephasing via the complementarity principle of electrons [5,6], the non-equilibrium decoherence [3, 4, 10, 11], and the temperature and magnetic field dependence of the dephasing have been investigated [1-4]. Due to the high visibility in MZI, which relies on the long coherence length in the chiral edge states of the electrons in the quantum Hall regime, we can quantitatively evaluate the coherence of the ballistic transport in solid state devices.

The noise in the electric current in mesoscopic devices serves as a powerful probe; the shot noise, which originates from the particle nature of electrons, clearly displays the characteristics of electron transport such as the charge of quasiparticles [12, 13] and many-body states of electrons [14-16], whereas the equilibrium noise (Johnson-Nyquist noise) is given by the product of the conductance and the temperature of the device. In electronic interferometers, it is discussed that the shot noise affects the electron coherence. For example, the shot noise plays a critical role in the controlled dephasing experiments due to the complementarity principle [5,6].

Here, we present an experimental study on the noise in a MZI. We mainly focus on the electron temperature dependence of the visibility; we first describe the result of the measurement of the equilibrium noise (Johnson-Nyquist noise) in the MZI, where we obtain reliable and quantitative information on the dephasing due to thermal fluctuations. In addition, we show the data of shot noise at MZI. The present results allow us to compare the dephasing due to the thermal fluctuations with the non-equilibrium noise quantitatively.

## 2. Experimental details

The MZI was fabricated with a GaAs/AlGaAs two-dimensional electron gas system (2DEG: electron density $2.0 \times 10^{11}$ cm$^{-2}$ and mobility $2.1 \times 10^6$

cm$^2$/Vs). In Fig. 1, the schematic geometry of the electric MZI is shown. The edge channels from the source Ohmic contact (S1) are injected to the MZI, which consists of two electronic beam splitters (quantum point contacts; QPC1 and QPC2, whose transmissions are tuned to be 50 % for the outer edge channel) and captured in the two drain contacts (D1 and D2). The current measured in D1 (or D2) oscillates as a function of the magnetic flux which perpendicularly penetrates the closed area formed by the two trajectories of the edge states. The area of the MZI estimated from the period of the Aharonov-Bohm (AB) oscillation is deduced to be around 14.3 μm$^2$. The visibility of the MZI ($v_I$), which measures the electron coherence, is defined as the contrast of the AB oscillation by the following conventional expression; $v_I = (I_{max} - I_{min})/(I_{max} + I_{min})$, where $I_{max}$ ($I_{min}$) represents the maximum (minimum) current of the oscillation.

The present experiment is performed in the IQHE regime at a filling factor (ν) of approximately $\nu$ = 1.66 (magnetic field perpendicular to the 2DEG (*B*) is 5 T). The AB flux is controlled by applying a voltage to the modulation gate (MG) to electrostatically deform the trajectory of the edge state as shown in Fig. 1. The QPC0 is used to select the edge channel for the interference experiment.

In this experiment, we tune the transmission of QPC0 to fully transmit the outer edge state while the inner channel is set to be fully reflected.

We measure the full current flowing from S1 to D2 by the standard lock-in technique at 31 Hz. The current fluctuations in D1 are measured by the noise measurement system shown in Fig. 1, which contains a cryogenic amplifier and a resonant circuit whose characteristic frequency is 3 MHz [17].

## 3. Results and Discussions

Figure 2 (a) shows a typical trace of the current measured at D2 as a function of the gate voltage applied to MG. This result is obtained at the optimized condition; the maximum visibility is about 50 % at the minimum electron temperature ($T_e$) at 20 mK and at $B$ = 5.0 T ($v$ = 1.66).

Figure 2 (b) shows the electron temperature estimated from the obtained Johnson-Nyquist noise. The current noise ($S_{th}$) is represented as $S_{th} = 4k_B T_e G$ at a zero-frequency limit, where $k_B$ is the Boltzmann constant, $T_e$ is the electron temperature, and $G$ is the conductance of the sample. According to this equation, we can extract the electron temperature by measuring $S_{th}$ and $G$ of the device [17].

The electron temperature dependence of the visibility [18-20] of the MZI at $B$ = 5 T is shown in Fig. 2 (c), with comparing to another result measured at 4.5 T ($v$ = 1.84). Above $T_e$ = 80 mK, decrease of the visibility can be expressed by $\exp(-T_e/T_0)$ with $T_0 \sim$ 30 mK for the data at $B$ = 5 T. It was discussed in Ref. [21] that $T_0$ is predominantly determined by the size of the interferometer and the present observation is consistent with Ref. [21] (see Fig. 3(a)). The visibility in the present case, however, seems to start to saturate at temperatures lower than 80 mK. If we naively extrapolate the observed exponential increase of the visibility to lower temperatures, the visibility would exceed 100 % at 0 K. Our direct measurement of the electron temperature substantiates the apparent saturation of the visibility at low temperatures reported in Ref. [3] and exclude electron overheating as the cause of this saturation. It is very interesting to note that the result well scales $(1+T_e/T_0)\exp(-T_e/T_0)$ for overall the temperature rather than a simple exponential function as shown in Fig. 2(c). Such temperature dependence was already discussed theoretically in Ref. [20].

In Fig. 3(b), the visibility of the MZI as a function of the bias voltage $V_{SD}$ is shown. This reproduces the well-known lobe-structure of mesoscopic interferometers [3, 10, 11, 21, 22]. Although the shape of the lobe structure is discussed to be an exponential function of the bias [22], we experimentally found that the lobe-structure is well fitted by the following empirical function [3, 11, 21]: $v_I = v_{I0}|\cos(\pi eV_{SD}/\varepsilon_L)|\exp[-(eV_{SD})^2/2\varepsilon_0^2]$, where $v_{I0}$ is the visibility at zero bias, and $\varepsilon_L$ and $\varepsilon_0$ are the characteristic energies; $\varepsilon_L$ is determined by the period of phase reversal and $\varepsilon_0$ is the characteristic width of the envelope, respectively. In the previous paper, we discussed these characteristic energies and the characteristic temperature $T_0$ in several interferometers with comparing the size and found universality between these parameters [21]. In Figs. 3(a) and (c), we plot these characteristic energies of the MZI obtained from the fitting in Figs. 2(c) and 3(b) and again found that the present values fall on our universal lines [23-25]. The interferometer size dependence of the visibility was discussed in Ref. [20] based on the electron-electron interaction in interferometers. The presence of such a universal relationship over various mesoscopic interferometers experimentally suggests further theoretical explanation for an underlying mechanism of dephasing.

By the direct comparison, the energy scales of dephasing estimated experimentally ($k_BT_0$, $\varepsilon_L$, and $\varepsilon_0$) are quite different from each other; $\varepsilon_L$, and $\varepsilon_0$ are about 10 times larger than $k_BT_0$ ($\varepsilon_L/k_BT_0 \cong 17$, $\varepsilon_0/k_BT_0 \cong 12$, in this experiment), although one may simply imagine that thermal fluctuation and the shot noise should have same energy scales. One of the possible explanations is intrachannel Coulomb interaction discussed by Youn *et al*; phase randomizing by the increase of non-equilibrium electrons was calculated in Ref. [26]. Another possible explanation for this difference was given by the theory by I. P. Levkivskyi *et al*. [19], which is based on the chiral Luttinger liquid approach to the quantum Hall state. Their theory predicts that $\varepsilon_L / k_B T_0 = 2\pi^2 \cong 19.6$, which might explain the present result quantitatively. Moreover, Roulleau *et al*. suggested recently that the coherence of the edge state is limited by the Johnson-Nyquist noise of the system [27]. In non-equilibrium system, the energy scale of the shot noise is suppressed by a factor $F$ (Fano factor: $0 \leq F \leq 1$) [14, 15, 28, 29] from the applied bias voltage; this shot noise reduction may explain the difference of the energy scale, while whether an interferometer dephases itself is an open question at this moment.

      Finally, we present the experimental result on the shot noise measurement. Figure 4 (a) shows the bias dependence of the visibility measured with the noise in D1. At this time, the maximum visibility at zero bias is about 26 % [30]. Figures 4 (b) and (c) display the voltage noise measured in D1

as a function of MG voltage with the oscillation pattern of the transmission to D2. At zero bias voltage (Fig. 4(b)), we observed clear oscillation of the voltage noise in D1 which perfectly follows the oscillation of the transmission (frequency: $f_0$) because the Johnson-Nyquist noise is proportional to the conductance of the device. On the other hand, at $V_{SD}$ = -30 µV, the correlation between the current and the noise decreases.

To evaluate the correlated component in Figs. 4(b) and (c), we performed Fourier transform to the noise (see Fig. 4(d)). One can see that there are two frequency components $f_0$ and $2f_0$ for the noise at $V_{SD}$ = -30 µV, while just $f_0$ at $V_{SD}$ = 0 µV. The $2f_0$ frequency component reflect the quadratic dependence of the transmission in the shot noise ($S_{shot}$); $S_{shot}$ = $2eIT_m(1-T_m)$, where $e$ is charge of an electron, $I$ is current, and $T_m$ is the transmission of the device. While the $f_0$ component of the noise is the sum of the thermal noise and the shot noise, this $2f_0$ oscillation directly is expected to show the coherence of the non-equilibrium current. Unfortunately, the amplitude of the noises estimated here are not quantitative enough. However, the relationship between the visibility and the amplitude of $2f_0$, $f_0$, and non-oscillating component of the noise will provide a more detailed understanding of the non-equilibrium transport through the interferometer.

## 4. Summary

In summary, we measured the equilibrium and non-equilibrium noise in the MZI. From the equilibrium noise, we obtained relevant information about the electron temperature dependence of the visibility. We proved that the visibility show saturation at very low electron temperature as shown in Ref. [3] by measuring the Johnson Nyquist noise of the MZI. We also measured the bias dependence and found the same universality in the relationship between the energy scales of the dephasing and the interferometer size as the previous report [21]. Moreover, we performed the measurement of the non-equilibrium noise in the MZI and found the second harmonic oscillation of the noise. The non-equilibrium noise will provide a sensitive probe for the statistics of charge transmission in mesoscopic interferometers.

This work is supported by KAKENHI, the Deutsche Forschungsgemeinschaft within the SFB631, Yamada Science Foundation, and Matsuo Science Foundation. MH thanks for the financial support by JSPS

**Figure 1**

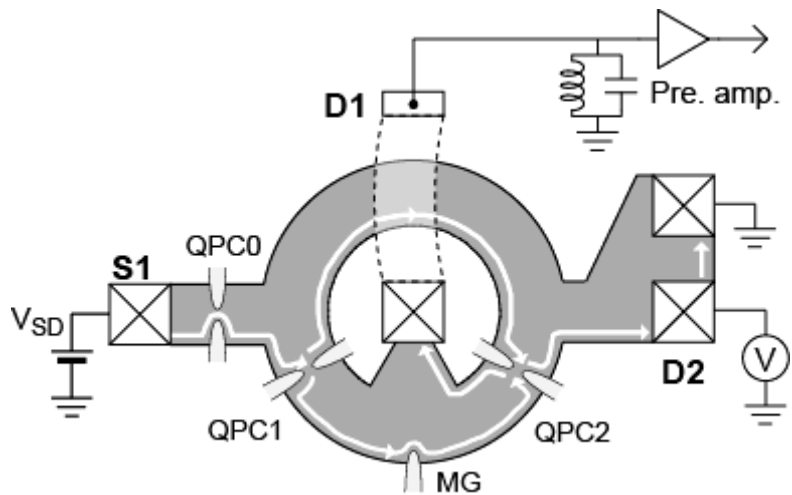

Fig. 1　Schematic illustration of the MZI sample and the measurement setup. Current derived from the source (S1) and drain (D2) are measured while (noise voltage noise) in D1.

**Figure 2**

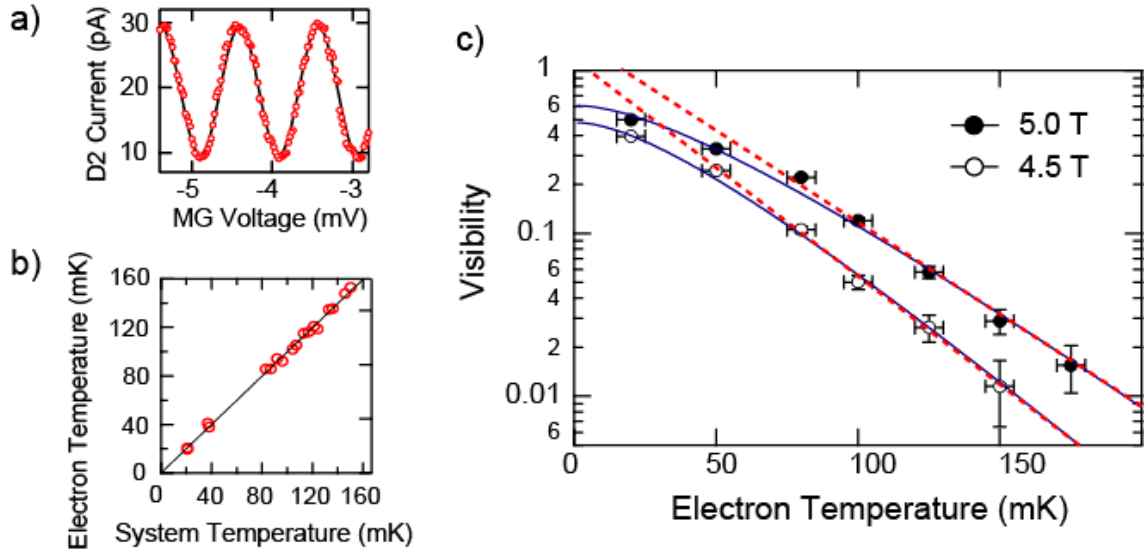

Fig. 2 (a) Aharonov-Bohm oscillation of current in D2 at optimized condition; 20 mK and 5 T. (b) Electron temperature estimated from the Johnson-Nyquist noise as a function of the system temperature. (c)Temperature dependence of visibility of the MZI. Red lines show the fitting curves by exponential functions. Blue lines show the guides for the experimental data by the functions of $(1+T_e/T_0)\exp(-T_e/T_0)$.

**Figure 3**

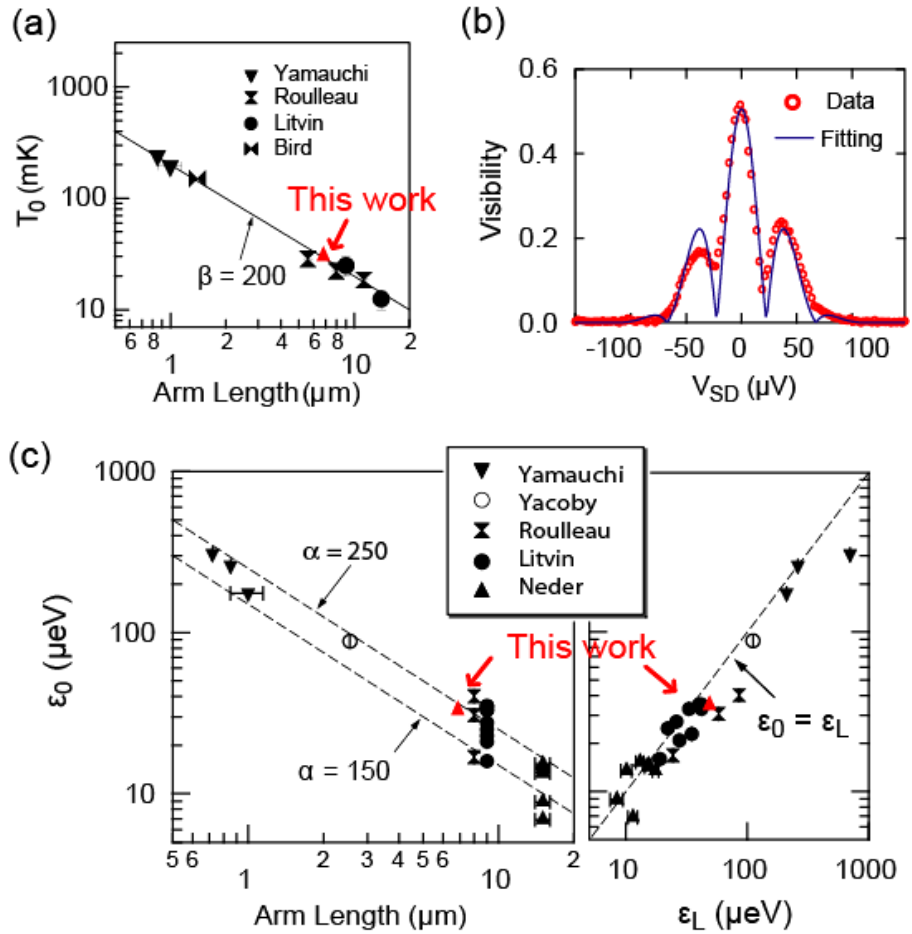

Fig. 3 (a) The dependence of $T_0$ on the arm length ($L$) of the interferometers is plotted by using the data obtained by the fitting in Fig. 2(c) and previous reports. These data are obtained around filling factor 2 [25]. (b) Visibility of the MZI as a function of $V_{SD}$. The solid line is the result of the fitting. (c) The energy scales of the lobe structure of the present MZI and the previous reports. Open symbols represent the data at zero magnetic field and filled symbols in the integer quantum Hall regime [25].

**Figure 4**

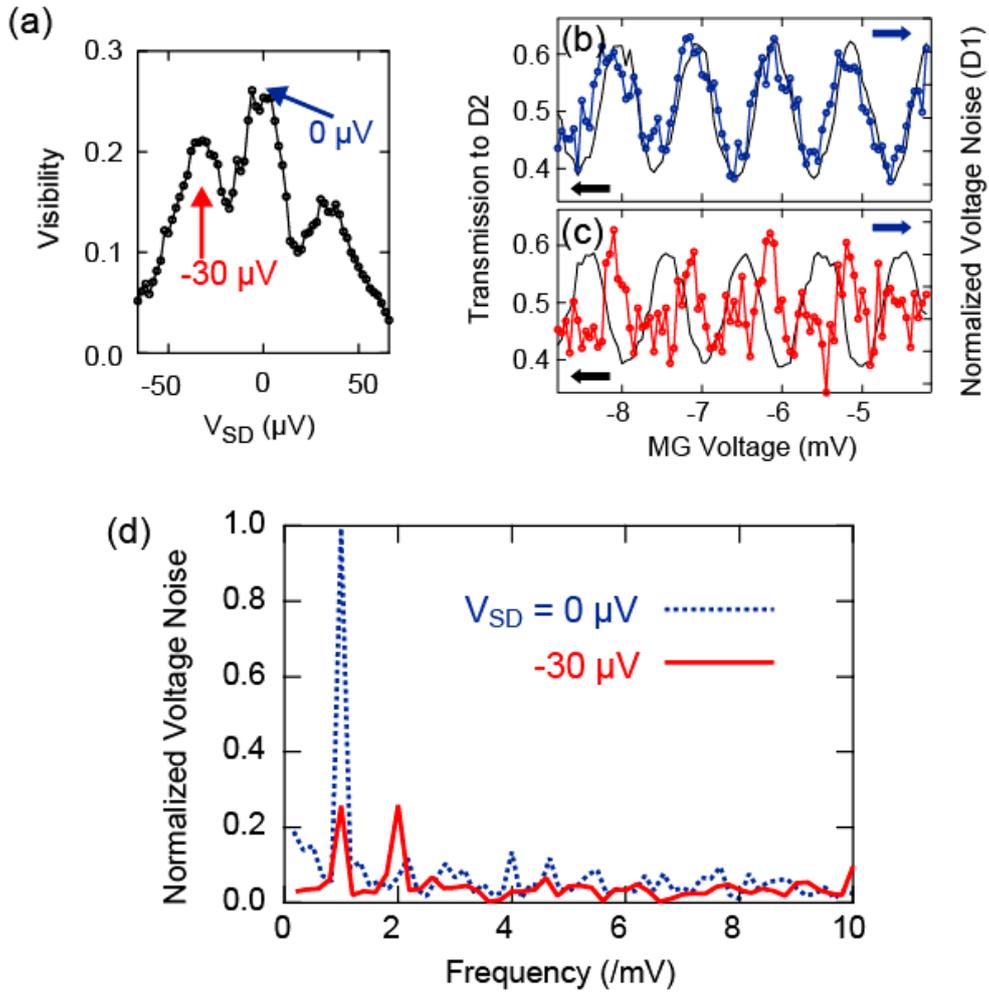

Fig. 4 (a) Lobe-structure of the MZI [30]. (b)(c) The normalized voltage noise in D1 as a function of the MG voltage with the transmission to D2. (b) and (c) were measured at $V_{SD}$ = 0 µV and $V_{SD}$ = -30 µV, respectively. (d) The result of the Fourier transform of the noise in Figs. (b)(c).